\RequirePackage{fix-cm}
\documentclass[smallextended]{svjour3}       
\smartqed  
\usepackage{graphicx}
\usepackage{braket}
\usepackage{pgfplots}
\usepackage{csquotes}
\usepackage{float}
\usepackage{multirow}
\usepackage{dcolumn}
\usepackage[colorlinks=true, citecolor=blue]{hyperref}
\usepackage{hhline}
\usepackage{amssymb}
\usepackage{amsmath}
\usepackage{graphicx}
\usepackage{subfigure}
\begin{document}

\title{Generation of perfect W-state and demonstration of its application to quantum information splitting
}


\author{Manoranjan Swain \and Vipin Devrari \and Amit Rai \and
        Bikash K. Behera \and Prasanta K. Panigrahi 
}


\institute{Manoranjan Swain \at
             Department of Physics and Astronomy,
              National Institute of Technology, Rourkela, 769008, Odisha, India \\       \email{swainmanoranjan333@gmail.com} 
             \and
              Vipin Devrari \at
              Department of Physics and Astronomy,
              National Institute of Technology, Rourkela, 769008, Odisha, India \\       
              \email{devrari.uk@gmail.com} 
              \and
              Amit Rai \at
              Department of Physics and Astronomy,
              National Institute of Technology, Rourkela, 769008, Odisha, India\\
            \email{amitrai007@gmail.com}           
           \and
           Bikash K. Behera \at
             Bikash's Quantum (OPC) Pvt. Ltd., Balindi, Mohanpur 741246, Nadia, West Bengal, India,   \at   Department of Physical Sciences, Indian Institute of Science Education and Research Kolkata, Mohanpur 741246, West Bengal, India \\   \email{bikash@bikashsquantum.com}   
              \and 
              Prasanta K. Panigrahi \at   Department of Physical Sciences, Indian Institute of Science Education and Research Kolkata, Mohanpur 741246, West Bengal, India \\ \email{pprasanta@iiserkol.ac.in} 
}

\date{Received: date / Accepted: date}

\maketitle

\begin{abstract}
We report the first experimental realization of perfect W-state in a superconducting qubit based system. In contrast to maximally entangled state, the perfect W state is different in weights and phases of the terms contained in the maximally entangled W-state. The prefect W state finds important applications in quantum information processing tasks such as perfect teleportation, superdense coding, secret sharing etc. The efficiency of generation is quantified by fidelity which is calculated by performing full quantum state tomography. To verify the presence of genuine nonlocality in the generated state, we experimentally perform Mermin's inequality tests. Further, we have also demonstrated splitting and sharing of quantum information using the experimentally generated state.
\end{abstract}

\section{Introduction}
Entanglement~\cite{l1,n1} has become the central part of quantum computation and quantum information processing~\cite{a4,a5} tasks such as quantum teleportation~\cite{m1,a3,m2,a2}, dense coding~\cite{m3,a25}, quantum cryptography~\cite{m4,m5,a17,a18,a19} etc. The correlation found between entangled quantum particles is quite different from that of classical correlation and are practically impossible according to classical physics~\cite{l1}. The entanglement of bipartite states has been extensively studied. However multipartite entanglement~\cite{p1,p2} which involves entanglement between more than two subsystems has a much more complicated structure. Multipartite entangled states find applications in quantum computing and quantum information processing tasks~\cite{a16,a15}. Moreover, they are also relevant in diverse area of physics such as condensed matter physics~\cite{a1} and quantum gravity~\cite{a114}. In the class of multi-partite entangled states, there exist the W-state~\cite{pw1} which is well recognized due to its robustness against particle loss. The general form of W-state~\cite{pnfl_1} is given as,
\begin{equation}\label{eq1}
\begin{split}
&\arrowvert W\rangle= \sum p_i |1_i,\{0\}\rangle,\\
\hspace{1cm} &\sum{|p_i|^2}=1.
\end{split}
\end{equation}
If one chooses $p_i$=$\frac{1}{\sqrt{N}}$(where N is the number of qubits), the obtained state is known as the maximally entangled W-state, which is given as,
\begin{equation}\label{eq3}
  \arrowvert W\rangle= \frac{1}{\sqrt{N}}(\arrowvert100.....0\rangle+ \arrowvert010.....0\rangle+........... + \arrowvert 0.....001\rangle)
\end{equation}

There are many proposals~\cite{gsa,pkp,a30,a31,a33} as well as experimental realizations~\cite{a32,a34,a35} of maximally entangled W-state. However, the usability of maximally entangled W-state in teleportation was questioned by Gorbachev $\emph{et al.}$~\cite{gorbachev} and was concluded to show non-unit fidelity in quantum teleportation by Joo  $\emph{et al.}$~\cite{pw2}. As suggested by Agrawal and Pati~\cite{pw3}, there exist another type of W-state which is useful for various quantum information processing tasks such as perfect quantum teleportation~\cite{pw3}, quantum superdense coding~\cite{pw3}, quantum information sharing~\cite{pw4}, quantum information splitting and sharing~\cite{pw5} etc. These states are known as perfect W-states. These are slightly different in weights and phases of the terms contained in the maximally entangled form of W-state. The generalized three-qubit perfect W-state is given as,

\begin{equation}\label{eq2}
  \arrowvert W_{p,s}\rangle= \frac{1}{\sqrt{2+2s}}(\arrowvert100\rangle+ \sqrt{s}e^{i\Phi_1}\arrowvert010\rangle+ \sqrt{s+1}e^{i\Phi_2}\arrowvert001\rangle)
\end{equation}
where s is a real number, $\Phi_1$ and $\Phi_2$ are phases acquired by the state. This general form contains a wide class of W-type states. For simplicity, we choose s=1, $\Phi_1$=0 and $\Phi_2$=0 and the resultant state can be represented as,
\begin{equation}\label{pw}
  \arrowvert W_{p,1}\rangle= \frac{1}{2}(\arrowvert100\rangle+ \arrowvert010\rangle+ \sqrt{2}\arrowvert001\rangle)
\end{equation}

The state represented in  Eq. $ \ref{pw}$ is found to have important applications in prefect teleportation and superdense coding~\cite{pw3}. So, the experimental generation of the state is essential for quantum computation and information processing tasks. 
Earlier various schemes were proposed to generate this state~\cite{pkp,pw6,pw8,pw9,pw10,pw12}. However, to the best of our knowledge, the experimental generation of such a state has not been reported yet. Here for the first time we present the experimental generation of the perfect W-state in a superconducting system based platform.

IBM, by developing a cloud based quantum computing platform~\cite{pw13} has created new opportunities for research in quantum physics. The building blocks of this platform are superconducting qubits which are maintained at a very low temperature to be in working condition. The coherence times of these qubits ranges from 5$\mu$s to 55$\mu$s. This system has been utilized to perform diverse experiments such as, experimental violation of inequalities in support of quantum entanglement~\cite{qv_f3,n1_Diego}, quantum router~\cite{bn1}, quantum simulations \cite{qv_LiertaarXiv2018,bn2}, demonstration of no-hiding theorem~\cite{l3}, demonstration of quantum secret sharing~\cite{l4}, demonstration of quantum walks~\cite{walk1,walk2,walk3} etc.

In the present work, we use IBM's one of the five qubit quantum computer to experimentally generate the three-qubit entangled perfect W-state represented in Eq. $\ref{pw}$. To show the quality of generation we have done full quantum state tomography of the obtained state and calculated the state fidelity. In addition to the first experimental generation, we have performed an application i.e. splitting of quantum information using the experimentally generated state.

Splitting of quantum information is a robust way of sending information as it requires mutual collaboration between the recipients. One can retrieve the information only when other recipients agree by becoming controller of the process. In this paper, we experimentally implement the scheme proposed in Ref. ~\cite{pw5} by considering perfect-W state as the quantum channel.

Our work in this paper is organized as follows. Section \ref{generation} and \ref{ent} give the details of experimental generation and entanglement verification of the state. In Section \ref{inf}, we have demonstrated splitting of quantum information. Finally we conclude in Section \ref{con}.

\section{Experimental generation of perfect W- state}\label{generation}
\noindent In this section, we give a detailed analysis on experimental generation of the perfect W-state.

To experimentally generate the state, we choose IBM's one of the superconducting qubit based five-qubit quantum computer platform, `ibmq-vigo'.

  \begin{figure}[h]
  \centering
  \includegraphics[width=0.9\textwidth,height=0.33\textwidth]{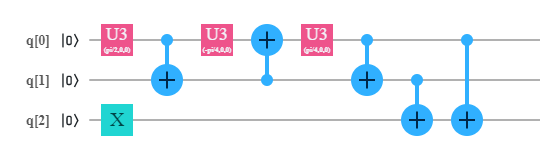}
  \caption{Circuit for generation of perfect W-state. X is the Pauli X-gate and U3($\theta$, $\phi$, $\lambda$) is the general unitary gate.}
  \label{circuit}
  \end{figure}

  \begin{figure}[h]
  \centering
  \includegraphics[width=0.6\textwidth,height=0.4\textwidth]{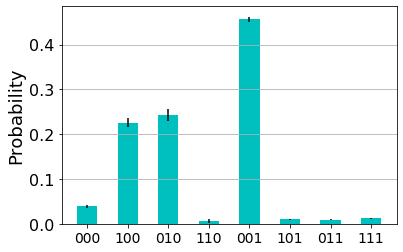}
  \caption{Experimental outcome of the circuit given in Fig.~\ref{circuit} when run for 8192 shots in ibmq-vigo. Bar represents the standard deviation from mean probability.  }
  \label{outcome}
  \end{figure}

  \begin{figure*}
  \centering
   \subfigure[]{\includegraphics[scale=0.33]{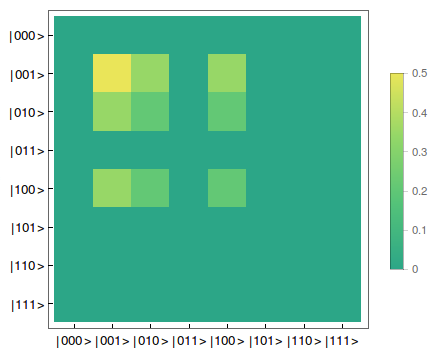}}\quad
  \subfigure[]{\includegraphics[scale=0.33]{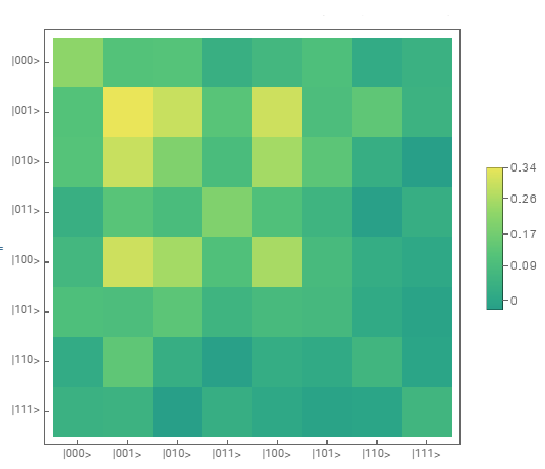}}\quad
  \subfigure[]{\includegraphics[scale=0.33]{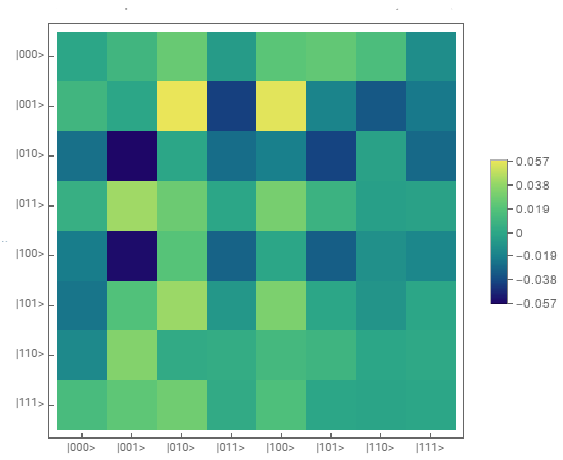}}
  \caption{(a) Theoretical density matrix with no imaginary values. (b) the real and (c) imaginary part of the experimental density matrix.}
  \label{density}
\end{figure*}


         
 %
         
        
         


 



The experimental result of the algorithm shown in Fig.~\ref{circuit} is presented in Fig.~\ref{outcome}. The result is to be followed by considering the first digit as the first qubit. For example in the term `100', the qubits are to be read in sequential manner i.e. `q0, q1, q2 $\to$ 1, 0, 0'.

To characterize the experimentally obtained state only $|Z\rangle$  basis measurements are not sufficient. To collect all information about the state complementary measurements are to be taken on the same state. The process  followed in this work to fully characterize the state is `Quantum state tomography' (QST)~\cite{pw31,bn3}. The objective of QST is to reconstruct the density matrix of the state by taking series of measurements on the obtained state. This reconstruction of density matrix needs $4^N$-1(N is the number of qubits) different measurements on the state. After the full quantum state tomography we followed Stokes's formula~\cite{pw31} to construct the density matrix. The stokes formula to calculate density matrix for three-qubit state is given by,

\begin{equation}\label{a1}
\hat{\rho} = \frac{1}{2^3}\sum_{i,j,k=0}^{3} (S_i \otimes S_j \otimes S_k)  \hat{\sigma}_i \otimes \hat{\sigma}_j \otimes \hat{\sigma}_k
\end{equation}

where $S_0$= $P_{|0\rangle}$ + $P_{|1\rangle}$,  $S_1$= $P_{|+\rangle_x}$ - $P_{|-\rangle_x}$,  $S_2$= $P_{|+\rangle_y}$ - $P_{|-\rangle_y}$ and  $S_3$= $P_{|0\rangle}$ - $P_{|1\rangle}$. $\sigma_0$, $\sigma_1$, $\sigma_2$ and $\sigma_3$ are identity and Pauli matrices respectively. The theoretical and experimental density matrices are represented in Fig.~\ref{density}. To check the closeness of the obtained state with the theoretical one we have calculated fidelity~\cite{pw32}, which is given as:
\begin{equation}\label{fid}
    F(\rho^T,\rho^E)= \bigg|Tr \bigg({\sqrt{\sqrt{\rho^T}\rho^E\sqrt{\rho^T}}}\bigg)\bigg|^2
\end{equation}
The fidelity of the state generation is calculated to be 0.75 $\pm$ 0.02.


\section{Non-locality of perfect W-state}\label{ent}
The non-locality of perfect W-state is verified by using Mermin's inequality~\cite{mermin}.

\begin{equation}\label{ineq}
  |M|=|E(ABC)-E(A B'C')-E(A'B'C)-E(A'BC')|\leq 2
\end{equation}

The three particle form of Mermin's inequality~\cite{merminform} represented in Eq. $ \ref{ineq}$. A, A$^\prime $, B, B$^\prime $, C, and C$^\prime $ are set of measurements on particles. Note that in quantum mechanics, these set of measurements can be considered as spin measurements and hence can be  specified by linear combination of Pauli spin operators. In the present case for the calculation of quantum mechanical expectation values of the terms appearing in the R.H.S of Eq. $\ref{ineq}$ the choice of spin measurements has been restricted to X-Z plane only. 


 Considering the measurement directions for unprimed ones to Z-direction and primed ones to X-direction, the value of Mermin's polynomial ($|M|$) is calculated to be $\approx$ 2.9.\\

To perform this task in IBMQ use choose ibmq-vigo. Following the process as has been done in~\cite{mp} the measurements are taken. The outputs of the measurements are listed in Table \ref{qv_TabI}.

\begin{table}[]
 \centering
  \caption{Outcome of the measurement results with their corresponding probabilities(P) $\pm$ standard deviation(SD) calculated from different sets of data each was run for 8192 shots on ibmq-vigo. By converting these probabilities to expectation values, Mermin's polynomial can be calculated.}
 \begin{tabular}{p{.7cm}| p{1.7cm}| p{1.7cm} |p{1.7cm}| p{1.7cm}}
 \multicolumn{3}{c}{} \\
 \hline
\hline
 OC & P $\pm$ SD for $ABC$ & P $\pm$ SD for $A'B'C$ & P $\pm$ SD for $A'BC'$ &P $\pm$ SD for $AB'C'$\\
 \hline
 \textbf{000}   & 0.039$\pm$0.003   & 0.253$\pm$0.007  & 0.370$\pm$0.018  & 0.370$\pm$0.007\\

         001    & 0.458$\pm$0.007   & 0.137$\pm$0.004  & 0.037$\pm$0.003  & 0.037$\pm$0.014\\

         010    & 0.235$\pm$0.007   & 0.020$\pm$0.002  & 0.059$\pm$0.012  & 0.039$\pm$0.009\\

 \textbf{011}   & 0.009$\pm$0.001   & 0.122$\pm$0.010  & 0.069$\pm$0.003  & 0.303$\pm$0.003\\

         100    & 0.230$\pm$0.003   & 0.025$\pm$0.005  & 0.055$\pm$0.012  & 0.065$\pm$0.003\\

 \textbf{101}   & 0.009$\pm$0.001   & 0.119$\pm$0.009 & 0.252$\pm$0.024  & 0.051$\pm$0.013\\

 \textbf{110}   & 0.004$\pm$0.002   & 0.235$\pm$0.004  & 0.090$\pm$0.006  & 0.083$\pm$0.009\\

         111    & 0.012$\pm$0.001   & 0.084$\pm$0.005  & 0.066$\pm$0.006  & 0.048$\pm$0.004\\
 \hline\hline
 \end{tabular}
\label{qv_TabI}
\end{table}
 \vspace{0.1cm}
 Using the data presented in Table \ref{qv_TabI}, the value of the Mermin's polynomial is calculated to be 2.516 $\pm$ 0.027. The result shows the violation of the inequality and hence verifies the non-locality of the obtained state.

\section{Splitting quantum information}\label{inf}
We start by giving a brief theoretical description of the process. Let us consider Alice, Bob and Charlie share an entangled perfect W-state of the form given below,
\begin{equation}\label{pw2}
  \arrowvert W\rangle= \frac{1}{2}(\arrowvert0_1 0_2 1_3\rangle+ \arrowvert0_1 1_2 0_3\rangle+ \sqrt{2}\arrowvert1_1 0_2 0_3\rangle)
\end{equation}

 Alice also possess an information qubit which is of the form,
\begin{equation}
    |\phi_0\rangle= \alpha|0_0\rangle+\beta|1_0\rangle
\end{equation}

The combined state is given as,

\begin{equation}\label{pwc}
\begin{aligned}
  \arrowvert \Psi \rangle= \frac{1}{2}\alpha(\arrowvert0_0 0_1 0_2 1_3\rangle+ \arrowvert0_0 0_1 1_2 0_3\rangle+ \sqrt{2}\arrowvert0_0 1_1 0_2 0_3\rangle) \\  + \frac{1}{2}\beta(\arrowvert1_0 0_1 0_2 1_3\rangle+ \arrowvert1_0 0_1 1_2 0_3\rangle+ \sqrt{2}\arrowvert1_0 1_1 0_2 0_3\rangle)
  \end{aligned}
\end{equation}
Alice then performs Bell measurement on her two qubits leaving Bob and Charlie as receivers. Neither Bob nor charlie can retrieve the information alone. Hence one of them has to become the controller and the other can retrieve the message by performing X and Z measurements on his qubit.

\begin{figure}
\centering
  \includegraphics[width=.6\linewidth]{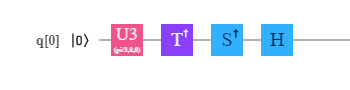}
 \caption{Arbitrary state generation circuit in ibmq-vigo.}
  \label{stckt}
\end{figure}

\begin{figure*}
\centering
\subfigure[]{\includegraphics[width=0.5\linewidth]{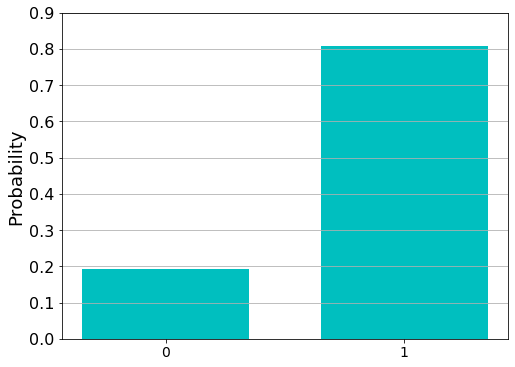}}\quad
\subfigure[]{ \includegraphics[width=0.5\linewidth]{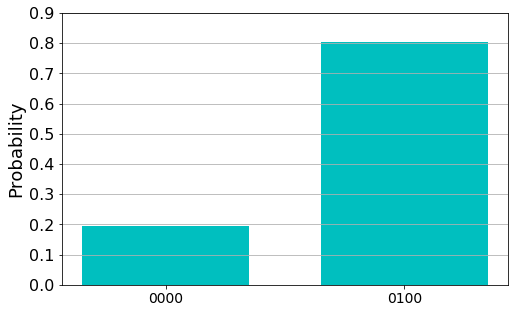}}\quad
 \subfigure[]{ \includegraphics[width=0.5\linewidth]{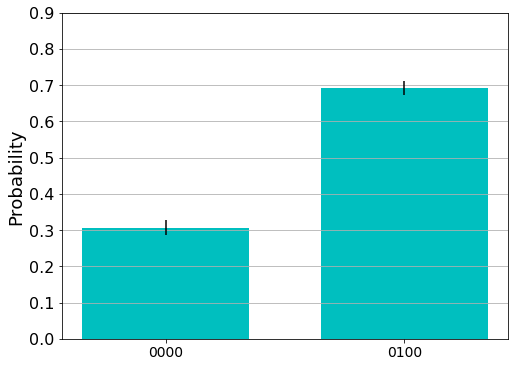}} \quad
 \caption{(a) State to be shared by Alice  (b) Simulation result of the circuit given in Fig.~\ref{stckt}, (c) Experimental result of communication.}
 \label{stateresult}
\end{figure*}

\emph{Experimental implementation -
}At initial stage, the qubits q1, q2, q3 are prepared in perfect W-state form. The qubit q0 which was in a state $|0\rangle$, is projected on to an arbitrary state by sequence of single qubit gates i.e. U3(pi/3,0,0), $T^\dagger$, $S^\dagger$, and H. The circuit to generate the state and its simulation result in ibmq-vigo are shown in  Fig.~\ref{stckt} and Fig.~\ref{stateresult}(a) respectively.

\begin{figure}
 \centering
 \includegraphics[width=0.9\textwidth,height=3.cm]{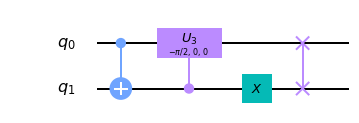}
  \caption{Unitary operation between Bob and charlie which makes Bob the receiver.}
\label{unitary}
\end{figure}

\begin{figure}
 \centering
 \includegraphics[width=1.2\textwidth,height=8cm]{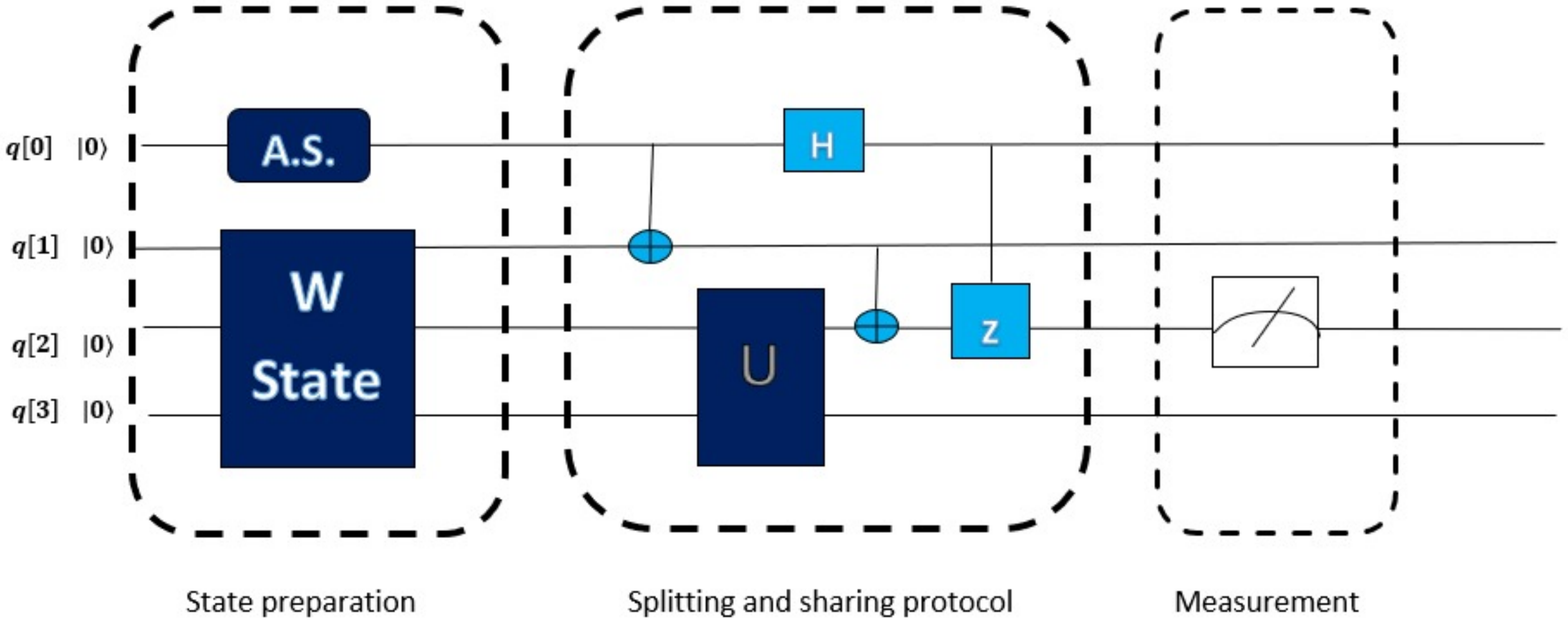}
  \caption{Quantum information splitting circuit. The first section is for generation of the entangled state. The second section is for the Bell basis measurement and the third section is for controlled operation and q2 measurement. A.S. is abbreviated for arbitrary state.}
\label{controlledcommunication}
\end{figure}

Now let us assume,  Alice has control over q0 and q1 and Bob and Charlie control q2 and q3 respectively. Alice performs Bell basis measurement on the two qubits she possess. The message can be received by any one from Bob and Charlie but not without the help of other. Let's suppose Charlie becomes the controller and Bob becomes the receiver in the present case. To achieve this process Bob has to perform two-particle unitary transformation on the qubits q2 and q3. The required unitary matrix is given as:

The circuit to implement this unitary operation between q2 and q3 is shown in Fig.~\ref{unitary}. This unitary operator projects charlie's qubit to state  $|0\rangle$ and the message signal is transferred to Bob. Finally Bob can retrieve the message by applying controlled Pauli X and Z gates on his qubit. The whole communication process is shown in Fig. ~\ref{controlledcommunication}.

The communication process circuit was simulated as well as experimentally run on ibmq-vigo device for 8192 shots. The simulation and experimental outcome of measurement on q2 is shown in Fig.~\ref{stateresult} (b) and (c) respectively. Following the process of quantum state tomography the experimental density matrix of the state was obtained. The theoretical and the experimental (one out of many sets of experimental data) density matrices of the state are represented by the following equations:

\begin{equation}\label{ext}
 \rho^T=
\begin{pmatrix}
0.194 & 0.250   \\ 0.250 & 0.806 
\end{pmatrix}
+
i
\begin{pmatrix}
0 & - 0.306 \\  0.306 & 0
\end{pmatrix}
\end{equation}

\begin{equation}\label{exd}
 \rho^E=
\begin{pmatrix}
0.289 & 0.174  \\ 0.174  & 0.709
\end{pmatrix}
+
i
\begin{pmatrix}
0 &  - 0.118  \\  0.118 & 0
\end{pmatrix}
\end{equation}

The fidelity of the communication process is calculated to be 0.805 $\pm$ .006.

\section{Conclusion}\label{con}
In conclusion we have shown the first experimental generation of perfect W-state on a superconducting qubit based platform with a fidelity of 0.75 $\pm$ 0.02. The Mermin inequality tests on the state clearly verified the presence of genuine entanglement between the qubits. Besides the experimental generation we have also shown controlled communication using the experimentally generated state. The fidelity of the communication process was obtained to be 0.805 $\pm$ .006.
\section*{Acknowledgments}
One of the authors (A.R.) gratefully acknowledges a research grant from Science and Engineering Research Board (SERB), Department of Science and Technology (DST), Government of India (Grant No. CRG/2019/005749) during this work. B.K.B. acknowledges the prestigious Prime Minister's Research Fellowship awarded by DST, Govt. India. Authors acknowledge IBM team for providing free access to their cloud computing platform.


%
%



\end{document}